# Lost in Transliteration: Bridging the Script Gap in Neural IR

Andreas Chari
a.chari.1@research.gla.ac.uk
University of Glasgow
Glasgow, United Kingdom

Iadh Ounis
Iadh.Ounis@glasgow.ac.uk
University of Glasgow
Glasgow, United Kingdom

Sean MacAvaney
Sean.MacAvaney@glasgow.ac.uk
University of Glasgow
Glasgow, United Kingdom

## Abstract

Most human languages use scripts other than the Latin alphabet. Search users in these languages often formulate their information needs in a transliterated –usually Latinized– form for ease of typing. For example, Greek speakers might use Greeklish, and Arabic speakers might use Arabizi. This paper shows that current search systems, including those that use multilingual dense embeddings such as BGE-M3, do not generalise to this setting, and their performance rapidly deteriorates when exposed to transliterated queries. This creates a "script gap" between the performance of the same queries when written in their native or transliterated form. We explore whether adapting the popular "translate-train" paradigm to transliterations can enhance the robustness of multilingual Information Retrieval (IR) methods and bridge the gap between native and transliterated scripts.[1] By exploring various combinations of non-Latin and Latinized query text for training, we investigate whether we can enhance the capacity of existing neural retrieval techniques and enable them to apply to this important setting. We show that by further fine-tuning IR models on an even mixture of native and Latinized text, they can perform this cross-script matching at nearly the same performance as when the query was formulated in the native script. Out-of-domain evaluation and further qualitative analysis show that transliterations can also cause queries to lose some of their nuances, motivating further research.

## CCS Concepts

• **Information systems** → **Information retrieval**.

## Keywords

Transliteration, Neural IR

## 1 Introduction

Human languages use a multitude of different scripts. For example, the Ethnologue language database reports over 7000 languages, of which 4,153 have a developed writing system.[2] Search users might, therefore, use a variety of scripts to express their information needs. Although many languages use an alphabet based on Latin characters, others use alphabets ranging from scripts similar to Latin, such as Cyrillic, to scripts completely different, such as Chinese (Hanzi) or Arabic. Transliteration is the process of converting text from one script to another. This is done to express the pronunciation of the original script in a different one whilst approximating the original pronunciation. For example, Ελληνική Δημοκρατία, i.e., "Hellenic Republic" in English, can be transliterated into the Latin script –or Romanized– as Helleniki Dimokratia. Humans use transliterations into scripts such as Latin for multiple reasons. For example, it is commonly used in second language learning for languages such as Arabic or Chinese to enable language learners to learn the pronunciation before learning to read and write in the native script [1]. Transliterations can also help non-speakers read a provided text, e.g., the name "Александр" can be written as "Aleksandr", allowing English speakers to pronounce it. Another reason people use transliteration is for ease of typing, especially with devices or software that do not accommodate non-Latin scripts, to avoid switching keyboards on smartphone devices, or to hide spelling mistakes [8, 22, 27].

In Information Retrieval (IR), the purpose of a search system is to satisfy a user's information need expressed in a query by typically retrieving relevant documents from a collection. This task is complicated for search users who use a transliterated version of their native language to express their information needs. Documents are commonly written in the native script of a language, whereas it can be easier for humans to express their information needs in a transliterated query. This adds further complexity in satisfying the users' needs when there is a "script gap" between the form of expression of their information needs in the query and the representation of the information in the document. To accommodate the needs of search users who might use transliterations, search systems not only need to be robust in languages other than English but also be able to close the "script gap", which is created when the search user uses transliterated queries. These queries, whilst technically in the same languages as the documents of interest, are expressed in a completely different script. While current search systems are typically not trained to bridge this cross-script gap, we aim to investigate if the use of neural methods might mitigate this due to the robustness of their text embedding or if they are also susceptible to the "script gap", thereby hindering the use of transliterated queries for neural retrieval.

In this paper, to investigate and bridge the script gap in multilingual IR, we answer the following research questions:
**RQ1**: How robust are existing neural IR methods when dealing with transliterated queries?

---
[1]https://github.com/andreaschari/transliterations

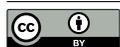

---
[2]https://www.ethnologue.com/faq/

**RQ2**: Can the script gap be bridged using "transliterate-train"?
**RQ3**: Does "transliterate-train" generalise to out-of-domain collections?

Our investigation focuses on the mMARCO dataset [3], specifically its Chinese and Russian translations. We use Uroman [9] to produce the respective queries' Latin transliterations (romanizations). First, we show that commonly used neural retrievers are not robust enough to match information across language scripts (RQ1). Even in languages such as Russian, whose Cyrillic script markedly overlaps with the Latin script, we observe a significant performance degradation. We investigate whether exposing the neural models in transliterated queries during fine-tuning would enable the model to close the "script gap" in performance between the transliterated and the native forms of the language. We adopt the commonly used "*translate-train*" [20] procedure for cross-script training (RQ2) and fine-tune the models on a combination of transliterated queries and native documents. We observe that fine-tuning on training data evenly split between native and transliterated queries enables the models to reduce the performance gap between the scripts and additionally report significant performance gains in native query performance. Furthermore, we investigate whether training the models only using transliterated queries will close the "script gap" even more. Our results show mixed outcomes depending on the language. Finally, we evaluate our models in an out-of-domain setting using the respective TREC NeuCLIR datasets for Russian and Chinese (RQ3). In the out-of-domain evaluation, we observe that whilst "transliterate-train" can generalise and close the script gap in Russian, it fails to do so for Chinese. Further qualitative analysis shows that transliterations cause the NeuCLIR queries to lose some nuances, leading to more generic results.

## 2 Related Work

Recent literature has seen considerable work in Natural Language Processing (NLP) on handling transliterations. Similarly, we have seen substantial work in neural IR, which aims to bridge language barriers with tasks such as cross-lingual and multilingual IR by leveraging neural techniques. While there has been research on transliterations in IR, there is a notable gap in investigating how to bridge transliteration barriers in *neural* IR. This section aims to cover the existing literature in each direction. In the transliteration literature, most work has focused on non-neural methods. Additionally, if neural methods are used, they are rarely used for an information retrieval downstream task. Prabhakar and Pal [23] looked at the recent transliteration literature and noted that as the number of non-English speakers among online search users increases, transliterated search will also likely become a growing phenomenon. Yadav et al. [30] studied various models of transliterations. Most approaches opted for statistical modelling rather than neural approaches, with only one prior work employing a BERT-based architecture [19]. The authors argued that hybrid approaches using neural networks will allow further development of robust transliteration models. Recent NLP work has also leveraged transliterations to enhance the multilingual capacity of neural models. Husain et al. [10] proposed using romanized text as a bridge representation to increase the support of low-resource languages that use non-Latin scripts. The authors observed that Romanized representations are more efficient compared to the native script ones and are better aligned with English representations. Similarly, Ma et al. [15] used transliterations to transfer LLMs to low-resource languages. Liu et al. [14] introduced a sequence-level pre-training objective to align various scripts using transliterations to address the script barrier. Similarly, Xhelili et al. [28] used transliterations to transfer multilingual pre-trained language models to low-resource languages that do not use a Latin script, such as Farsi and Korean. Their method aligns various scripts at the token and sentence levels during post-training.

Transliterated queries were the object of study in the Transliterated Search and Mixed Script IR Shared Tasks at the Forum for Information Retrieval Evaluation (FIRE) conference [2, 6, 25, 26]. Their task was constrained by the state of IR development at that time. From the available information, we observed that most submissions opted for methods based on BM25 and TF-IDF. Indeed, we would expect modern search techniques that use pre-trained language models to be more robust than these methods.

Pre-trained language models have allowed IR research to enhance the robustness of search systems by leveraging their linguistic capabilities for retrieval. Recent years have seen the use of pre-trained language models pretrained on a large number of non-English languages, for example, XLM-R [7] and mT5 [29] as neural rankers. Recent work, such as Chen et al. [5], leveraged XLM-R to produce a multilingual embedding model called BGE-M3. mT5 [29], a multilingual variation of T5 [24], pre-trained on a dataset covering 101 languages has been later used by Bonifacio et al. [3] for retrieval by fine-tuning mT5 on mMARCO. Recent works, such as Chari et al. [4], have begun investigating the effects of different surface forms of queries on these neural IR methods. However, their study is limited to differences in spellings between British and American English. To the best of our knowledge, there has not been any investigation on the effects of transliterating the entire query to a different script in neural IR.

The "translate-train" paradigm is commonly used to train cross-lingual IR systems. In Nair et al. [20], *translate-train* was used to train their ColBERT-X model on pairs of the English queries and translated documents in Chinese, Persian and Russian. Similarly, Lawrie et al. [13] applied "translate-train" to develop an efficient method for multilingual IR. The performance of methods leveraging the "translate-train" paradigm motivates us to extend it to a "cross-script" setting where the query and document pairs share the same languages. In our task, the queries are machine-translated into a Romanized script rather than a different language.

Recent literature has also seen the development of non-English multilingual corpora for training and evaluation. The mMARCO [3] collection is a popular resource containing automatically translated versions of MSMARCO [21] in more than ten languages. The TREC NeuCLIR track [11, 12] released NeuCLIR/1, a collection of Persian, Chinese and Russian Common Crawl News documents as well as neuMARCO, a cross-language automatically translated version of MSMARCO. Both collections are now standard benchmarks for evaluating the cross-language capabilities of search systems.

While all of these resources are excellent and have merit, none currently benchmark the robustness of IR methods when dealing with transliterated queries, opting for either a monolingual or a cross-lingual evaluation. Indeed, the current search evaluation paradigm is insufficient for scenarios where the query and the document share the same language but are expressed using a different script. Hence, our paper will aim to address this gap.

Table 1: Performance of BGE-M3 and mT5 models on Chinese and Russian using mMARCO/v2

| Model \Queries | Chinese (N) | | Chinese (T) | | Russian (N) | | Russian (T) | |
| --- | --- | --- | --- | --- | --- | --- | --- | --- |
| | MRR@10 | R@1000 | MRR@10 | R@1000 | MRR@10 | R@1000 | MRR@10 | R@1000 |
| BGE-M3 | 0.2342 | 0.8940 | 0.0078 | 0.0550 | 0.2444 | 0.8998 | 0.1244 | 0.6589 |
| BGE-M3 (N) | 0.2732† | 0.9125† | 0.0071 | 0.0524 | 0.2838† | 0.9137† | 0.1238 | 0.5925† |
| BGE-M3 (50) | 0.2642† | 0.8991 | 0.1382† | 0.6807† | 0.2770† | 0.9117† | 0.2633† | 0.8895† |
| BGE-M3 (T) | 0.2112† | 0.7967† | 0.1608† | 0.7255† | 0.2785† | 0.9105† | 0.2679† | 0.8963† |
| BGE-M3 >> mT5 | 0.2441 | 0.8940 | 0.0081 | 0.0550 | 0.2552 | 0.8998 | 0.1948 | 0.6589 |
| BGE-M3 (N) >> mT5 (N) | 0.2503 | 0.9125† | 0.0070 | 0.0524 | 0.2522 | 0.9137† | 0.1192† | 0.5925† |
| BGE-M3 (50) >> mT5 (50) | 0.2520† | 0.8991 | 0.1105† | 0.6807† | 0.2533 | 0.9117† | 0.2339† | 0.8895† |
| BGE-M3 (T) >> mT5 (T) | 0.2070† | 0.7967† | 0.1145† | 0.7255† | 0.2662† | 0.9105† | 0.2522† | 0.8963† |

## 3 Transliterate-Train

We leverage a variation of the "translate-train" approach [20] to overcome data sparsity when training cross-script neural rankers. In "translate-train", a large dataset of good-quality relevance judgements, such as MSMARCO [21], is modified by applying automatic machine translation, resulting in a large dataset for other languages, e.g., the mMARCO collection [3]. For our task, we add a further step and produce automatic transliterations (Romanizations) of the training queries to leverage the strengths of these datasets and close the *script gap* between the native and transliterated queries. We denote this approach as "transliterate-train". This method exposes the neural rankers to queries and documents in the same language, albeit expressed using different forms of representation – i.e., different scripts. Therefore, the models are trained in pairs of either transliterated or native queries and native documents. This enables the rankers to be effective when the user uses a transliterated query, thereby yielding relevant documents.

## 4 Experimental Setup & Results

In our experiments, we use the Chinese and Russian translations of mMARCO/v2 [3] and neuCLIR/1 [12]. We focus on Chinese and Russian because they represent two completely different cases of languages that can be transliterated. Chinese characters have no overlap with Latin, and Russian's Cyrillic script overlaps by a large margin with the Latin script. Using Uroman [9], a tool commonly used to romanize non-Latin scripts, we produce romanized versions of both the 'dev/small' and the 'train' query sets for both languages. Finally, we repeat the process for the 'trec-2022' and 'trec-2023' query sets for the respective neuCLIR datasets. In our experiments, we focus on the single-vector representation component of BGE-M3 [5] as the retriever and mT5 [29] as the reranker. For retrieval and evaluation, we use the PyTerrier platform [18] and ir-measures [16]. We report MRR@10 and R@1000 for all experiments with mMARCO and nDCG@20 and R@1000 for neuCLIR. We use the ir_datasets [17] package for accessing the datasets. We fine-tune BGE-M3 and mT5 in three configurations, changing the number of transliterated queries in the training data while constantly using the native script documents. For our first configuration (N), we fine-tune the neural rankers only on the native-text queries. For our mixed configuration (50), we fine-tuned on an even split of transliterated and native text queries. Finally, for our third configuration (T), we fine-tune the neural rankers only on the transliterated queries. We perform significance testing with Bonferroni correction between each model configuration and their respective baselines, "BGE-M3" and "mT5" for the same query set. We use the † symbol to denote significant results.

### 4.1 RQ1: How robust are neural IR methods when dealing with transliterated queries?

Table 1 shows that multilingual neural IR methods cannot effectively handle the transliterated queries. For both Chinese and Russian, we observe significant degradations in the performance of BGE-M3 with a 97% and 49% drop in MRR@10, respectively. Even for Russian, whose Cyrillic script substantially overlaps with Latin, a decrease of almost half of the MRR performance is noteworthy, demonstrating how even minor differences in scripts can cause significant performance degradation. We observe similar patterns with the mT5 reranker, which shows that even the cross-encoder models are susceptible to this script gap. In preliminary experiments, we investigated if whitespace in the Chinese tokenisation is a factor for the low performance. We find that tokenising the queries before transliteration [3] yields inconsistent results between mT5 and BGE-M3. We therefore leave the study of tokenisation for future work. Overall, in answering RQ1, we observe that multilingual IR methods *are not* robust to transliterated queries.

### 4.2 RQ2: Can the script gap be bridged using "transliterate-train"?

Table 1 shows the results of our "transliterate-train" configurations. We observed even more significant gains for BGE-M3 during our preliminary experiments on the Chinese transliterated queries with the mixed configuration (50) when we pre-tokenised the queries before fine-tuning. However, the results did not hold for mT5. Hence, we leave the investigation of the impact of tokenisation in cross-script neural IR for future work and focus on the non-tokenised experiments for the rest of this paper. Examining BGE-M3, we observe that the native configuration (N), as expected, yields models with even more robust performance on the native queries. This does not transfer to any gains for the transliterated queries, with the Chinese models' performance remaining essentially the same. At the same time, we observe performance losses compared to the Russian baseline. Hence, we conclude that we cannot use the native configuration and fine-tune the models only on the original script to close the script gap. We can make the following observations from our experimental results for the mixed configuration (50). For BGE-M3, we observe significant improvements in the performance of transliterated queries in both languages. Similarly, mT5 also benefits from this fine-tuning configuration. We observe that the performance of the Russian queries is almost identical regardless of the script. In contrast, there remains a considerable, albeit reduced, difference between the native and transliterated queries for Chinese. This can be due to the similarities between Cyrillic and Latin scripts, making matching the scripts a much easier task for the models than matching Latin with Chinese. We also make the following observations for our experiments for the transliterated configuration (T). This configuration yields the best performance on the transliterated queries for BGE-M3 in both languages. Compared to the mixed configuration, we see more substantial performance gains in Chinese than in Russian transliterated queries. We observe the opposite behaviour on mT5, with the Chinese performance barely improving while noting gains in the Russian transliterated queries. Looking at the native queries, the transliterated configuration results in performance drops only for the Chinese queries. On the other hand, we observe that the performance of the Russian native queries remains broadly the same for BGE-M3, while we even note some modest gains for mT5. We can now answer RQ2 with the following concluding remarks. In general, bridging the gap between the transliterated and native Chinese scripts is a considerably more challenging task than for Russian. Overall, the mixed configuration is the only fine-tuning configuration, which preserves the native text performance and significantly enhances the robustness of both models on the transliterated queries (50). Finally, our experiments show that *we can* bridge the gap between the transliterated and native scripts on the query side using our proposed "transliterate-train" procedure.

---

[3]We used Pyterrier's Anserini interface: https://github.com/seanmacavaney/pyterrier-anserini/tree/main/pyterrier_anserini

Table 2: Performance of BGE-M3 and mT5 on Chinese and Russian using the combined neuCLIR trec 2022 and 2023 queries.

| Model \Queries | Chinese (N) | | Chinese (T) | | Russian (N) | | Russian (T) | |
| --- | --- | --- | --- | --- | --- | --- | --- | --- |
| | nDCG@20 | R@1000 | nDCG@20 | R@1000 | nDCG@20 | R@1000 | nDCG@20 | R@1000 |
| BGE-M3 | 0.4279 | 0.7500 | 0.0028 | 0.0251 | 0.4373 | 0.7460 | 0.2031 | 0.5018 |
| BGE-M3 (50) | 0.3886† | 0.6466† | 0.0698† | 0.2488† | 0.3924† | 0.6403† | 0.3293† | 0.6070† |
| BGE-M3 >>mT5 | 0.4237 | 0.7500 | 0.0028 | 0.0251 | 0.4418 | 0.7460 | 0.2970 | 0.5018 |
| BGE-M3 (50) >>mT5 (50) | 0.3380† | 0.6466† | 0.0605† | 0.2488† | 0.3541† | 0.6503† | 0.2851 | 0.6070† |

### 4.3 RQ3: Does "transliterate-train" generalise to out-of-domain collections?

Table 2 shows the results of our out-of-domain evaluation on neuCLIR. Given that only the mixed configuration bridged the script gap and retained the native script performance, we only focus on this configuration for the out-of-domain evaluation. For Russian, we observe that the mixed configuration enables significant gains in transliterated performance. We also observe considerable drops in the native script performance. For Chinese, we observe drops in the native script performance and modest gains in the performance of the transliterated queries. We also perform a qualitative analysis to investigate why our models fail to generalise consistently in the out-of-domain evaluation setting. We sample ten queries from the "trec-2022" query set and compare the top ten retrieved documents for both the transliterated and native script versions of each query. We observe that the transliterations cause queries to lose their nuance, leading to the retrieval of less relevant documents. For example, the query with "qid" 32: "花生過敏的治療" (Treatment of peanut allergy) is transliterated to "huashengguomindezhiliao". The transliterated query retrieves mostly generic documents about medicine, whereas the native script query retrieves documents primarily about allergies. Likewise, the query with "qid" 26: "乌克兰总统候选人泽连斯基" (Ukrainian presidential candidate Zelensky) is transliterated to "wukelanzongtonghouxuanrenzeliansiji". While the top ten documents retrieved with the native script focus exclusively on the presidential campaign, the transliterated query drifts and leads to generic documents about Zelensky or his presidential term after the election. Most of the ten queries sampled have only three or fewer common documents across scripts. This leads us to conclude that the transliterations create ambiguities, resulting in a decrease in effectiveness when dealing with the more technical documents of neuCLIR compared to the MMARCO documents on which we trained the models.

## 5 Conclusions

We investigated the robustness of multilingual neural IR methods when dealing with transliterated queries. Our experiments show that competitive neural IR models like BGE-M3 and mT5 are not robust to this scenario. We demonstrate that their performance is indeed severely affected when dealing with transliterated queries, even in languages such as Russian, whose Cyrillic script markedly overlaps with the Latin script. We experiment with a "transliterate-train" paradigm and demonstrate that by further fine-tuning the neural methods on a mixture of native and transliterated queries, we can significantly improve their robustness when dealing with transliterated queries. Further evaluation on out-of-domain datasets and a qualitative analysis show that transliterations can create ambiguities around the nuances of queries. Therefore, further research is needed on the proper transliteration procedure for neural IR. We hope this work re-invigorates the investigation of neural IR methods when dealing with other languages that commonly use transliterations or other mixed-script scenarios.